\documentclass[aps,prb,twocolumn]{revtex4-1}

\usepackage{t1enc}
\usepackage[utf8]{inputenc}
\usepackage{amsmath}
\usepackage{graphicx}

\begin{document}

\title{From stochastic single atomic switch to nanoscale resistive memory
device}

\author{A. Geresdi}
\author{A. Halbritter}
\author{A. Gyenis}
\author{P. Makk}
\author{G. Mihály}
\affiliation{Department of Physics, Budapest University of
Technology and Economics and \\
Condensed Matter Research Group of the Hungarian Academy of Sciences, 1111
Budapest, Budafoki ut 8., Hungary}
\date{\today}

\maketitle

\noindent\textbf{We study the switching characteristics of nanoscale junctions
created between a tungsten tip and a silver film covered by a thin ionic
conductor layer. Atomic-sized junctions show spectacular current induced
switching characteristics, but both the magnitude of the switching voltage and
the direction of the switching vary randomly for different junctions. In
contrast, somewhat larger junctions with diameters of a few nanometers display a
well defined, reproducible switching behavior attributed to the formation and
destruction of nanoscale channels in the ionic conductor surface layer. Our
results define a lower size limit of \(3\,\)nm for reliable ionic nano-switches,
which is well below the resolution of recent lithographic techniques.}

Solid state ionic conductors are good candidates for the next generation of
nonvolatile computer memory
elements\cite{nature03190,1375000,nature06932,nmat2023,nmat2748}. Such devices
have to show reproducible resistance switching at reasonable voltage and current
values even if scaled down to the nanometer sizes. Solid state memory blocks
based on resistive switching have already proven to operate down to
the lateral resolution of \(100\,\textrm{nm} - 1\,\mu\)m
(Refs.~\citenum{nature03190}, \citenum{nmat2023}, \citenum{Yang2008},
\citenum{4140579}, \citenum{nl073224p}). A further decrease
of the size, down to the ultimate limit of the single atomic diameter
\cite{0953-8984-22-13-133001}, raises many technical challenges including
reproducibility, which is a basic requirement for mass-scale production
\cite{1375000,5257331}. In such atomic-scale devices the possible junction
configurations are determined by the atomic granularity of matter and the fine
details of interatomic interactions. The transport through these nanostructures
with dimensions comparable to the Fermi wavelength is governed by the quantum
nature of conductance: the current is carried along discrete conductance
channels. The conductance of a junction is described by the Landauer
formula \cite{Landauer_form}:  \(G=G_0 \sum_i \tau _i\), the universal
conductance quantum \(G_0=2e^2/h \approx \left( 12.9\,k\Omega
\right)^{-1}\) being multiplied by the sum of the channel transmission
probabilities determined by the details of the junction geometry and the
electronic structure of the material \cite{Scheer1998}.

\begin{figure}
\includegraphics[width=8.25cm]{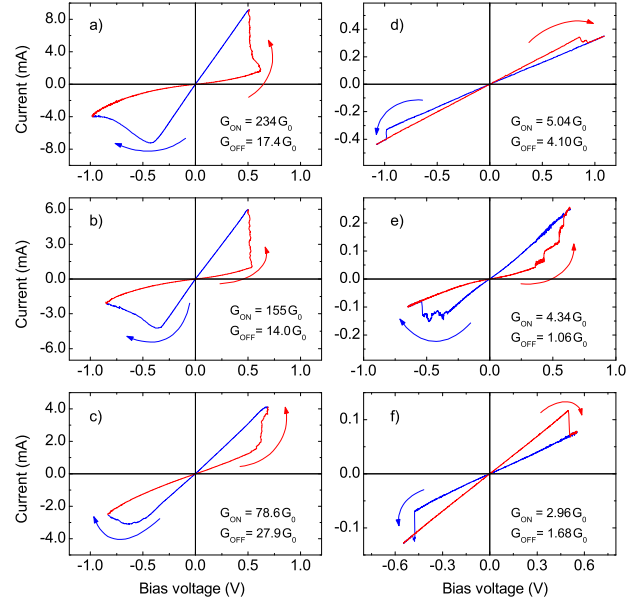}
\caption{\textit{Resistive switching phenomena for various selected
point contacts with different conductance values. The corresponding on-
and off-state conductance values measured at \(V=0.1\,\)V are shown on each graph,
respectively. The voltage bias is regarded as positive (negative) if the
Ag sample is positively (negatively) biased with respect to the W tip.}}
\label{Fig1}
\end{figure}

In this work we study the resistive switching phenomenon on variable size
nanoscale junctions with diameters below a few tens of nanometers. Variable size
point contacts were created by gently touching an Ag thin film sample with an
electrochemically sharpened W tip. To enhance the stability of the system, the
junctions were created and the measurements were performed at liquid He
temperature. More details on the experimental techniques are given in the
supplementary material.

In order to statistically characterize the junctions over a broad
scale of diameters we acquired   \(\approx 10^4\) I-V curves by
creating contacts with different conductance values
(\(1\,G_0\,-\,400\,G_0\)). The diameters of the junctions, estimated by
the Sharvin expression \cite{Sharvin}, vary from single atom
size to \(\approx 10\,\)nm.

In this regime the I-V curves regularly
exhibit a clear reversible switching behavior in the voltage
range of \(0.1-1\,\)V (more than \(60\%\) of the junctions have shown
jumps with relative amplitudes above \(10\%\)).
Figure \ref{Fig1} shows typical I-V characteristics for contacts of various
conductances.  Both the current and voltage values seem to be
surprisingly high for a device with dimensions of a few nanometers. The
corresponding current densities exceed \(10^9\,\)A/cm\(^2\). Note,
however, that in this mesoscopic transport regime the contacts are not destroyed
by the Joule heat, as the dissipation occurs inside a much larger volume
determined by the inelastic scattering length \cite{PhysRevB.62.9962,
Agrait200381}.

\begin{figure}
\includegraphics[width=8.25cm]{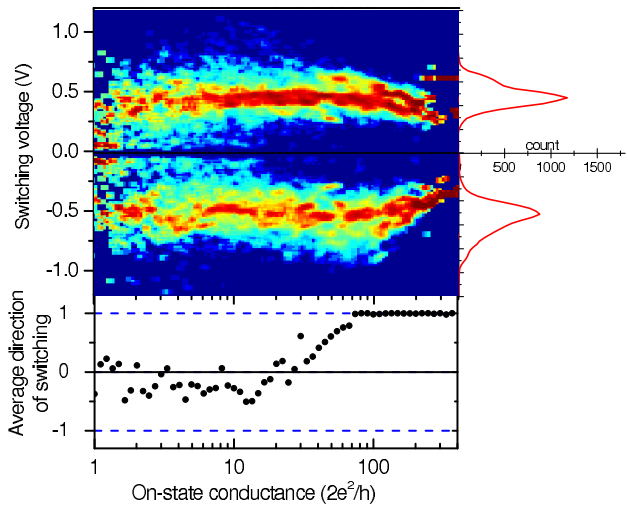} \caption{\textit{(a)
Switching voltage density plot for I-V curves of different on-state conductance
values between \(1\,G_0\) and \(300\,G_0\). Color scale represents the
relative occurrence of switching threshold voltage as a function of
\(G_\textrm{ON}\).
(b) Switching voltage histograms for curves with
\(G_\textrm{ON}>50\,G_0\). (c) The averaged switching direction as a
function of \(G_\textrm{ON}\). \(+1/-1\) represents well defined direction
described in the text, while \(0\) means random direction.}} \label{Fig2}
\end{figure}

A careful comparison of the curves shown in Fig.~\ref{Fig1} reveals significant
differences between the characteristics of the junctions with typical on-state
conductance of \(G_\textrm{ON}=50-300\,G_0\) (panels on the left) and that of the
smaller contacts with on-state conductance below about \(G_\textrm{ON}=20\,G_0\)
(panels on the right). Junctions in the first, higher conductance regime will be
denoted as \emph{nanoscale junctions}. For such contacts, the current steeply
increases as soon as a critical voltage is exceeded. The conductance, however,
does not saturate at a predefined value, it increases as long as the applied
voltage is ramped up. In our system a serial resistance of \(R_s=90\,\Omega\) is
present, which limits this process as the junction's resistance becomes
comparable to \(R_s\). Note, that in this regime the circuit is not purely
voltage driven, which accounts for the \emph{back-turning} of the I-V curve in
Fig.~\ref{Fig1} (a). We found, that in these \emph{nanoscale junctions} the
direction of the current-voltage loops is the same for all the curves.

In contrast, smaller junctions with
\(G_\textrm{ON} < 20\,G_0\) -- denoted as \emph{atomic-sized contacts}
-- exhibit a switching behavior between different well defined
initial and final states. During the switching event the system jumps to a new
state, corresponding to a new line in the I-V characteristics, and stays there
both for increasing and decreasing voltage ramps (until a back-switching occurs
for reversed voltage). The switching voltage, however, scatters in a broad range.
Moreover, the direction of the loops is random.

\begin{figure}
\includegraphics[width=8.25cm]{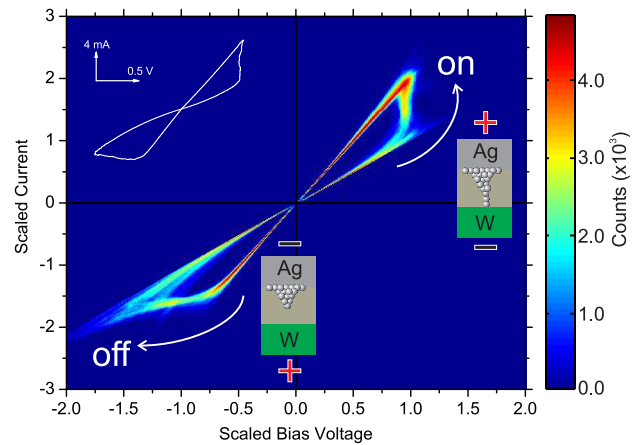} \caption{\textit{Color
density plot of 3500 independent I-V curves normalized to fixed
zero bias conductance values. Details of the scaling are given in the
supplementary material. The color scale represents the probability of the scaled
I-V values. The inset shows a typical switching characteristic.}}
\label{Fig3}
\end{figure}

The above features, presented on selected curves, are confirmed by statistical
analysis of a large amount of junctions. In Fig.~\ref{Fig2} (a) the
switching voltage values are shown in the form of a 2D colorscale plot. It is
evident that for \emph{nanoscale junctions} (\(G_\textrm{ON} > 50\,G_0\))
the switching voltage has a relatively well defined value at both polarities
around \(0.5\,\)V, whereas for \emph{atomic-sized contacts} (\(G_\textrm{ON} <
20\,G_0\)) the switching voltage scatters randomly in a broad voltage region. An
even more obvious distinction is observed in the \emph{direction} of the
switching. This is represented by \(+1/-1\) value  if the bias voltage of
the sample is positive/negative with respect to the tip for the
\mbox{off\(\rightarrow\)on} transition. Averaging the
direction of the switching for various junctions with similar conductance, one
finds that the switching is always positive for \(G_\textrm{ON}>50\,G_0\),
whereas for \(G_\textrm{ON}<20\,G_0\) the sign of the switching varies randomly
(Fig.~\ref{Fig2} (c)). We argue that the two
distinguished switching mechanisms are related to fundamentally different
physical phenomena. First we discuss the reproducible resistive switching
observed in highly conducting contacts, and later we return to the
random switching observed at the atomic level.

The curves recorded for \emph{nanoscale junctions} are very similar to those
recently reported for resistive switches based on Ag-Ag$_2$S-Me sandwich
structures\cite{terabe:4009, nature03190, 0957-4484-20-9-095710, jz900375a},
where Me is a transition metal and Ag$_2$S is an ionic conductor layer formed
between the two electrodes. The operation of these devices is based on the
electrochemical reaction \mbox{\(Ag_{(Ag_2S)}^+ + e^- \rightleftharpoons
Ag_{(metal)}\)} that is, the solubility of \(Ag^+\) ions in the Ag$_2$S lattice
allows migration controlled by the bias voltage applied. For asymmetric contacts,
reversible metallic filaments are built up and destructed by applying a
negative/positive voltage, respectively, which is reflected by abrupt changes in
the device conductance. The ionic migration is an activated
process\cite{jz900375a} resulting in a threshold voltage bias. In our
measurements the sign of the switching, the value of the threshold voltage and
the overall shape of the curves are consistent with earlier
experimental\cite{terabe:4009, nature03190, 0957-4484-20-9-095710, jz900375a} and
theoretical\cite{PhysRevLett.92.178302} reports on such ionic switches. We
emphasize, however, that in our measurements no special treatment of the sample
was performed, exposing the Ag layer to air for more than one week was enough to
establish the ionic-type switching behavior. More details on sample
characterization are provided in the supplementary material.

In order to demonstrate the universal and reproducible nature of this phenomenon
we scaled the individual I-V curves recorded in the range from \(50\,G_0\) to
\(400\,G_0\) by normalizing them to fixed, \(G_\textrm{OFF}=1\) and
\(G_\textrm{ON}=2\) zero bias conductance values. The resulting color density map
-- acquired out of more than \(3500\) independent I-V curves -- is shown in
Fig.~\ref{Fig3} demonstrating that the character of the I-V curve is very similar
for a broad variety of \emph{nanoscale junctions}. It is notable that the branch
of lower conductance exhibits more pronounced nonlinearity compared to the upper
one indicating the tunneling nature \cite{simmons:1793} of the low-conducting
state, which is shunted when metallic filaments are formed. This distinction is
even more clearly visible in Fig.~\ref{Fig1} (a), (b) and (c). The different
characters of the \mbox{on\(\rightarrow\)off} and \mbox{off\(\rightarrow\)on}
transition also show clear similarities to the results of earlier measurements
\cite{0957-4484-20-9-095710, nmat2023, C0NR00298D}.

\begin{figure}
\includegraphics[width=8.25cm]{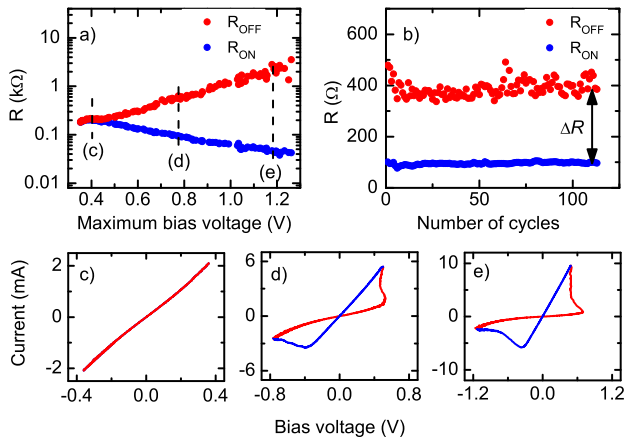}
\caption{\textit{(a) The evolution of the on- and off-state zero bias conductances as a
 function of the amplitude of the voltage ramp for a nanoscale ionic
 switch. Panels (c,d,e) show the I-V curves at three different
 bias voltage amplitudes indicated in panel (a). Panel (b)
 shows the on- and off-state conductances of a switching device during 100
 repeated cycles with a constant amplitude of \(0.75\,V\) of the voltage
 ramp. All of these measurements were performed on the same junction.}}
\label{Fig4}
\end{figure}

It was also studied how the states of the ionic switch can be tuned by the
applied bias voltage. Figure \ref{Fig4} (a) demonstrates that \emph{both} states
can be altered considerably. For this particular junction, the onset of the
resistive switching is at \(0.45\,\)V. By gradually increasing the amplitude of
the voltage ramp the on- and off-state resistance values are rapidly tuned. The
larger the resistance ratio, the larger threshold is observed for the
\mbox{off\(\rightarrow\)on} transition, which we attribute to a wider insulating
gap formed between the metallic sample and the tip.

We found that for constant amplitude ramping the switching phenomenon shows
excellent reproducibility. In this stationer state no apparent change is
observed during more than \(100\) cycles, as shown in Fig.~\ref{Fig4} (b). This
stability and reproducibility is characteristic of the ionic switching.
Decreasing the contact diameter below about \(3\,\)nm, however, results in a
crossover to a fundamentally different switching phenomenon.

\begin{figure}
\includegraphics[width=8.25cm]{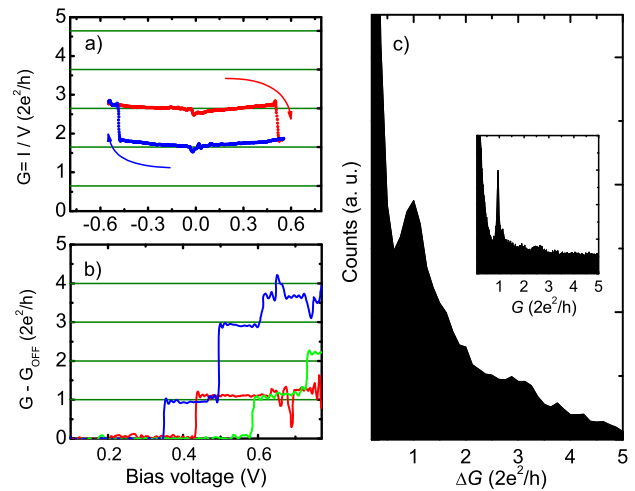} \caption{\textit{(a)
Conductance versus bias voltage curve indicating atomic-scale conductance
switching. Green lines are guides to the eye showing a periodicity with \(\Delta
G=1\,G_0\) starting from the low-conductance state. (b) Details of the
\mbox{off\(\rightarrow\)on} switching for three independent conductance versus
bias voltage curves with the off-state conductance subtracted. (c) Histogram of
the conductance change during the voltage sweep. \(\Delta G\) is calculated with
respect to the zero bias value of the conductance. The histogram is created for
130 independent I-V curves with \(G_\textrm{ON}<10\,G_0\). The inset shows the
histogram of the conductance versus electrode separation curves recorded during
5000 repeated opening and closing cycles at a constant bias voltage of
\(100\,\)mV.}}
\label{Fig5}
\end{figure}

In case of the \emph{atomic-sized contacts} the qualitative observations,
especially the random switching sign and the broad range of the threshold
voltages, contradict the picture of ionic switching. In this regime we attribute
the switching process to atomic rearrangements in the junction due to
electromigration induced by the high current density. The relevance of single
atomic displacements is demonstrated in Fig.~\ref{Fig5} (a) and (b), where
typical switching traces are presented as conductance-voltage curves. Since the
conductance value of a single Ag atom corresponds to the conductance quantum,
\(1\,G_0\) (Ref.~\citenum{Agrait200381}), jumps of this magnitude indicate the
inclusion or
 removal of a single atom in the junction area. Multiple
steps are often observed with amplitudes close to the integer multiplies of
\(G_0\) further supporting this scheme.

The role of the single atomic migration is also supported by the statistical
analysis of a large number of switching characteristics. A histogram was made of
the conductance traces detected during the voltage sweeps by subtracting the
baseline conductance measured in a narrow window (\(50\,\)mV) around zero bias.
The result is shown in Fig.~\ref{Fig5} (c). The well defined peak around
\(1\,G_0\) reflects that relative jumps corresponding to single atom conductance
are dominating. For comparison, the inset of Fig.~\ref{Fig5} shows the
traditional conductance histogram measured on the same sample by repeatedly
indenting the tip to the surface and breaking the such created nanojunction at a
constant bias voltage of \(100\,\)mV.

In a control experiment the above phenomenon was investigated in the same voltage
and current ranges utilizing the mechanically controllable break junction
technique (MCBJ), where the junction is created by the in-situ breaking of a
silver wire \cite{Agrait200381}. The freshly broken surfaces created in cryogenic
vacuum warrant the absence of ionic contamination in the junction. This method
fully reproduced the main features of the switching characteristics observed for
atomic scale contacts [Fig.~\ref{Fig1} (d)-(f)], while those of larger junctions
were not observed. Preliminary results obtained by the MCBJ technique on pure Au,
Pt, Fe revealed that the current induced atomic switching is a quite general
phenomenon in atomic-sized junctions.

In conclusion, we have identified two resistive switching mechanisms at the
nanoscale. It was shown that the ionic conductor-based devices are good
candidates for nonvolatile memory cells as they exhibit stable and reproducible
switching behavior. It was demonstrated, that solely the exposure of an Ag thin
layer to air is enough to establish an ionic conductor surface layer, which is
sufficient to form a reliable switching device in a nanoscale point-contact
geometry. Our results set lower limit for the size of such memory cells, below
which the character of the switching process dramatically changes due to the
enhanced single atomic migration. This size limit is estimated to be
 \(d \approx 3\,\)nm, resulting in storage densities above the capacity of
 current NAND Flash devices \cite{5257331} and comparable to the proposed bit
 size of magnetic media determined by the superparamagnetic limit \cite{824418}.

The authors are grateful to L.~Bujdosó and E.~Szilágyi for sample preparation and
characterization. This work was supported by the New Hungary Development Plan
under project ID: TÁMOP-4.2.1/B-09/1/KMR-2010-0002 and by the Hungarian Research
Funds OTKA under grants No. 72916 and No. 76010. A.~H.~is a grantee of the Bolyai
János scholarship.


%

\end{document}